\documentclass[10pt,oneside,onecolumn]{elsart}

\usepackage[ansinew]{inputenc}
\usepackage[english]{babel}
\usepackage{ifthen,shortvrb}
\usepackage[fleqn]{amsmath}
\usepackage{amsxtra,amsfonts,latexsym,amssymb,euscript}
\usepackage{amstext}
\usepackage{graphicx}
\usepackage{dcolumn}
\usepackage{subfigure}
\usepackage{arydshln}

\begin{document}

\begin{frontmatter}
\title{A coupled cohesive zone model for transient analysis of thermoelastic interface debonding}
\author{Alberto Sapora} and \author{Marco Paggi\corauthref{mp}}
\address{Department of Structural, Geotechnical and Building Engineering, Politecnico di Torino, Corso Duca degli Abruzzi 24, 10129 Torino}
\corauth[mp]{Corresponding author. Tel: +39-011-090-4910, Fax:
+39-011-090-4899, Email: marco.paggi@polito.it}

\begin{abstract}
A coupled cohesive zone model based on an analogy between fracture
and contact mechanics is proposed to investigate debonding phenomena
at imperfect interfaces due to thermomechanical loading and thermal
fields in bodies with cohesive cracks. Traction-displacement and
heat flux-temperature relations are theoretically derived and
numerically implemented in the finite element method. In the
proposed formulation, the interface conductivity is a function of
the normal gap, generalizing the Kapitza constant resistance model
to partial decohesion effects. The case of a centered interface in a
bimaterial component subjected to thermal loads is used as a test
problem. The analysis focuses on the time evolution of the
displacement and temperature fields during the transient regime
before debonding, an issue not yet investigated in the literature.
The solution of the nonlinear numerical problem is gained via an
implicit scheme both in space and in time. The proposed model is
finally applied to a case study in photovoltaics where the evolution
of the thermoelastic fields inside a defective solar cell is
predicted.\\
\vspace{1em} \noindent Note: this is the author's version of a work
that was accepted for publication in Computational Mechanics.
Changes resulting from the publishing process, such as editing,
structural formatting, and other quality control mechanisms may not
be reflected in this document. A definitive version was published in
Computational Mechanics, Vol. 53, October 2013, 845-857,
DOI:10.1007/s00466-013-0934-8
\end{abstract}

\end{frontmatter}

\section{Introduction}
\label{intro} The problem of stress and heat transfer across an
interface between elastically dissimilar materials is relevant in
engineering applications. If the bodies are initially separated and
then pressed into contact, surface roughness is a limiting factor to
achieve the conductivity of the bulk. A strategy to take into
account the effect of roughness in finite element computations has
been proposed in \cite{WZ1} by implementing a modified penalty
formulation with a contact law based on a thermo-plastic microscopic
contact model \cite{SY} in the node-to-segment contact geometry. In
case of geometrically linear problems where finite displacements in
the contact region do not take place, a simplification of the
rigorous formulation in \cite{WZ1} by using two-node contact
elements has been proposed in \cite{WZ}. Since the adopted physical
law contains dependencies from variables whose values change during
the analysis, an elegant consistent linearization of the
constitutive equations was proposed for the nonlinear iterative
procedure.

Another set of problems where stress and heat transfers across an
interface have to be computed is when initially fully bonded bodies
progressively debond in tension due to thermomechanical
deformations. The constitutive relations have to characterize the
progressive reduction of stress transfer and heat flux due to
increasing interfacial damage. In the framework on nonlinear
fracture mechanics, a thermomechanical cohesive zone model for
bridged delamination cracks in laminated composites has been
proposed in \cite{hatt,hatt2}. This thermomechanical cohesive zone
model formulation has been revisited in \cite{ozde} and an
application to polycrystalline materials under Mixed Mode
deformation was presented. In building physics, the interface
conductivity of bonded joints is an important property for the
assessment of reliability of insulation by using the well-known
Glaser diagram. In this field, a coupled problem between the thermal
field and the moisture diffusion can be of interest to avoid
humidity condensation inside insulated walls. In certain cases,
coupling with the elastic field has to be considered to predict the
occurrence of plaster decohesion. In this class of problems,
interface cracks require specific constitutive models to depict
decohesion, moisture and heat transfer. This led in \cite{moonen} to
a hygro-thermomechanical cohesive zone model specific for modelling
the phenomena of conduction in porous media. Other computational
work in this area regards hygro-mechanical problems at interfaces
\cite{1,2,3}, a coupled problem which shares some features with
thermomechanics.

In the aforementioned contributions related to the thermomechanical
behaviour of interfaces, either in compression or in tension, the
heat flux is considered to be dependent on the interface closure
(for contact mechanics) or opening (for fracture mechanics).
However, there are several applications in the field on
nanocomposites \cite{shah} where a constant interface conductivity
is used. This approach, called Kapitza model, can be regarded as a
constant spring in the framework of nonlinear spring elements. In
spite of its simplicity, this approach permits to simulate a range
of interface behaviors from highly conductive to perfectly
insulated, depending on the value of the Kapitza resistance.
Examples restricted to the thermal problem without coupling with the
mechanical one are discussed in \cite{LQ,Y}. Although the
mathematical formulation is simpler due to the lack of coupling
between the elastic and the thermal variables in the interface
constitutive relation, the Kapitza coefficient is hard to be
identified unless the interface is a well defined intermediate
material region with a given thickness.

In this study, a novel thermomechanical cohesive zone model is
proposed for the study of decohesion at material interfaces due to
thermal and mechanical loads. As compared to the state-of-the-art
literature on this matter, several novelties are presented. The
interface contact conductivity relation and its coupling with the
crack opening is derived by exploiting an analogy with contact
mechanics of rough surfaces, using the recent results established in
\cite{barb,paba}. This leads to an interface constitutive relation
with a limited number of free parameters of physical meaning that
can be identified from the quantitative analysis of roughness of
cracked interfaces. Moreover, the thermal analysis focuses on the
transient regime, obtained according to a solution strategy implicit
both in space and in time. Previous studies were limited to the
analysis of the steady state solution. A comparison between the
proposed approach with a fully coupled heat-conduction model
dependent on the displacement field and the uncoupled formulation
based on the Kapitza model is proposed. Contrary to the contact
problem in \cite{WZ}, where the coupling term was found to be of low
importance for the considered example, in the present case the
unsymmetrical coupled term of the stiffness matrix is relevant due
to the nonlinearity of the thermoelastic cohesive zone model.
Finally, an application to photovoltaics is proposed to show the
effect of cohesive cracks on the thermoelastic fields inside a
defective solar cell.

\section{Formulation of the thermomechanical problem with cohesive interfaces}\label{FEM}

\begin{figure}
\centering
  \includegraphics[width=0.5\textwidth]{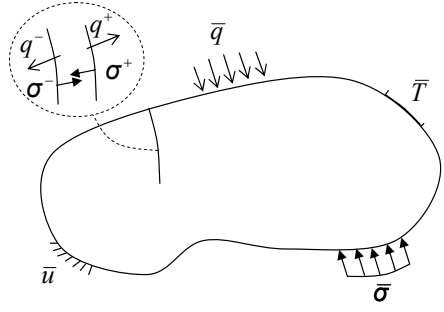}
\caption{Solid body with a cohesive interface.} \label{patata}
\end{figure}

The partial differential equations governing the mechanical
equilibrium in a solid body (Fig.\ref{patata}) with volume $V$ and
surface $S$ written in vectorial form are:
\begin{equation}\label{heat2}
\nabla^{\mathrm{T}} \textbf{S} \textbf{+f=0},
\end{equation}
where $\nabla$ is the gradient vector, $\textbf{S}$ is the Cauchy
stress tensor and $\textbf{f}$ is the vector of body forces. By
introducing the displacement vector $\textbf{w}$ and the stress
vector $\sigma$, the weak form corresponding to Eq. \eqref{heat2},
i.e. the principle of virtual works, writes:
\begin{equation} \label{weak2}
\begin{array}{ll}
 & \displaystyle{\int_V \textbf{S} : \nabla (\delta \textbf{w}) \mathrm{d}V =}\\
& \displaystyle{\int_V \textbf{f}^{\mathrm{T}} \delta \textbf{w}
\mathrm{d}V+\int_S \overline{\sigma}^{\mathrm{T}} \delta \textbf{w}
\mathrm{d}S+ \int_{S_{int}} \sigma^{\mathrm{T}} \delta(\Delta
\textbf{w})\mathrm{d}S,}
\end{array}
\end{equation}
where $\overline{\sigma}$ is the vector of prescribed tractions on
the boundary, while $S_{int}$ represents the internal surface. Note
that in Eq. \eqref{weak2} the notation
$$\textbf{S} : \nabla (\delta \textbf{w})=S_{ij}\dfrac{\partial \delta w_i}{\partial x_j} $$
has been adopted.

The partial differential equation governing the transient heat
conduction problem in the solid reads:
\begin{equation}\label{heat1}
-\nabla^{\mathrm{T}}\textbf{q}+Q=d_m\,c\,\dot{T},
\end{equation}
where $\textbf{q}$ is the heat flux vector, $Q$ is the heat
generation per unit volume per unit time, $d_m$ is the material
density, $c$ is the specific heat and $T$ is the temperature. By
means of Fourier's law $\textbf{q}=-k \nabla T$, $k$ being the
material conductivity, Eq.\eqref{heat1} can be rewritten as:
\begin{equation}\label{heat1:2}
k \nabla^2 T+Q=d_m\,c\,\dot{T}.
\end{equation}
The weak form related to the heat conduction problem, i.e. the
variational form of the energy balance, is then expressed as:
\begin{equation} \label{weak1}
\begin{array}{ll}
&\displaystyle{\int_V \textbf{q}^{\mathrm{T}} \nabla(\delta T) \mathrm{d}V =- \int_V k (\nabla T)^{\mathrm{T}} \nabla \delta T \mathrm{d}V=}\\
&\displaystyle{\int_V (d_m\,c\,\dot{T}-Q)\delta T\,
\mathrm{d}V+\int_S \overline{q} \delta T\, \mathrm{d}S+
\int_{S_{int}}q \delta (\Delta T)\,\mathrm{d}S.}
\end{array}
\end{equation}
where $\overline{q}$ represents the prescribed external heat flux
per unit area, normal to the boundary.

The last terms in Eqs.\eqref{weak2} and \eqref{weak1} represent the
contribution of the cohesive tractions and heat flux for the
displacement jump, $\Delta \textbf{w}$, and temperature jump,
$\Delta \textbf{T}$, across the interface.

\section{Finite element discretization of thermoelastic cohesive interfaces}

\begin{figure}
\centering
\includegraphics[width=0.6\textwidth]{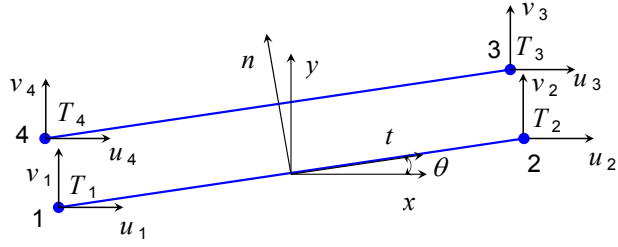}
\caption{Bi-dimensional linear interface element.}
\label{interfaccia}
\end{figure}

The coupled thermomechanical problem for the continuum can be
discretized by using standard four-node quadrilateral finite
elements (FE) with a mixed formulation. As regards the cohesive
interfaces, a four-node linear interface element compatible with the
elements used to discretize the continuum can be introduced, as
sketched in Fig.\ref{interfaccia}. As compared to the 2D formulation
for mechanical problems \cite{S93,pawr2}, each node has three
generalized degrees of freedom in the global reference system
instead of two: the horizontal displacement $u_i$, the vertical
displacement $v_i$ and the temperature $T_i$. In 3D, four degrees of
freedom for each node have to be specified. These generalized
displacements can be collected in the element vector $\mathbf{u}$:
\begin{equation}\label{eq1}
\mathbf{u}=\left(u_1,v_1,T_1,u_2,v_2,T_2,u_3,v_3,T_3,u_4,v_4,T_4\right)^T.
\end{equation}

A local reference system defined by the tangential vector $t$ and
the normal vector $n$ to the interface element is introduced, see
Fig.\ref{interfaccia}. The origin $O$ of the local reference system
is placed in the center of the element, which is in general rotated
with respect to the global $x-$axis by an angle $\theta$. The
generalized vector $\textbf{u}^*$ of the $i-$th node in the local
coordinate system can be computed via a pre-multiplication by a
rotation matrix, $\mathbf{u_i}^*=\mathbf{r}\mathbf{u_i}$, i.e.:
\begin{equation}\label{eq2}
\left\{
\begin{array}{c}
u_i^* \\
v_i^* \\
T_i^*
\end{array}
\right\}= \left[
\begin{array}{rrr}
  \cos\theta & \,\sin\theta & 0 \\
  -\sin\theta & \,\cos\theta & 0 \\
  0 & 0 & 1
\end{array}
\right] \left\{
\begin{array}{c}
u_i \\
v_i \\
T_i
\end{array}
\right\}.
\end{equation}
Therefore, the generalized displacement vector of the whole
interface element in the local reference system, $\mathbf{u}^*$, can
be related to $\mathbf{u}$ as follows:
\begin{equation}\label{eq3}
\mathbf{u}^*=\mathbf{R}\mathbf{u},
\end{equation}
where $\mathbf{R}$ is obtained by the collection of the individual
rotation matrices $\mathbf{r}$:
\begin{equation}\label{eq4}
\mathbf{R}=\left[
\begin{array}{cccc}
 \mathbf{r}  & 0 & 0 & 0 \\
 0  & \mathbf{r} & 0 &  0\\
 0  & 0 & \mathbf{r} & 0 \\
 0  & 0 & 0 & \mathbf{r}
\end{array}
\right].
\end{equation}
The relative generalized displacement vector $\mathbf{\Delta
u}^*=(u_4^*-u_1^*,v_4^*-v_1^*,T_4^*-T_1^*,u_3^*-u_2^*,v_3^*-v_2^*,T_3^*-T_2^*)^{\mathrm{T}}$
can now be computed as $\mathbf{\Delta u}^*=\mathbf{L}\mathbf{u}^*$,
where the operator matrix $\mathbf{L}$ relates the displacement and
temperature field components to the relative displacements and
temperatures between the upper and the lower sides of the interface,
$\Gamma^+$ and $\Gamma^-$:
\begin{equation}\label{eq5}
\mathbf{L}=\left[
\begin{array}{cccccccccccc}
 -1  & 0  & 0  &  0 & 0  & 0 & 0  & 0 & 0 & +1 & 0  & 0\\
 0   & -1 & 0  &  0 & 0  & 0 & 0  & 0 & 0 &  0 & +1 & 0\\
 0   & 0  & -1 &  0 & 0  & 0 & 0  & 0 & 0 &  0 & 0  & +1\\\
 0   & 0  & 0  & -1 & 0  & 0 & +1 & 0 & 0 &  0  & 0 & 0\\
 0   & 0  & 0  & 0  & -1 & 0 &  0 & +1 & 0 & 0 & 0 & 0\\
 0   & 0  & 0  & 0  &  0 &-1 & 0 &   0 &+1 & 0 & 0 & 0\\
\end{array}
\right].
\end{equation}
The vector of the tangential, normal and temperature gaps for a
generic point along the interface element,
$\mathbf{g}=(g_{\mathrm{t}}, g_{\mathrm{n}}, g_T)^{{\mathrm{T}}}$,
can be determined from $\mathbf{\Delta u}^*$ using standard
interpolation functions, $\mathbf{g}=\mathbf{N}\mathbf{\Delta u}^*$,
where $\mathbf{N}$ is given by
\begin{equation}\label{eq6}
\mathbf{N}=\left[
\begin{array}{cccccc}
 N_1  & 0  & 0 & N_2  & 0 & 0  \\
 0    & N_1 & 0  & 0 & N_2 & 0 \\
 0 & 0 & N_1 &  0 & 0 & N_2\\
\end{array}
\right].
\end{equation}
In the present case, $N_1=(1-s)/2$ and $N_2=(1+s)/2$ are the linear
shape functions. The $s$-coordinate ranges between $-1$ and $+1$, as
for standard two-node isoparametric finite elements.

The vector $\mathbf{g}$ can therefore be related to the nodal
generalized displacement vector as follows:
\begin{equation}\label{eq7}
\mathbf{g}=\mathbf{N}\,\mathbf{L}\,\mathbf{R}\,\mathbf{u}=\mathbf{B}\,\mathbf{R}\,\mathbf{u}.
\end{equation}
At this point, the constitutive relations for the interface,
relating tractions and heat flux to displacement and temperature
gaps, have to be introduced. For the sake of generality, we consider
now a whatever nonlinear relation between those quantities. In the
next section, a specific model will be introduced and the equations
particularized to that case. The contribution to the weak form by
the interface elements (Eqs. \eqref{weak2} and \eqref{weak1})
written in a compact way is:
\begin{equation}\label{Gint:1}
\delta G_{int}=\int_{S_{int}} \delta \textbf{g}^{\mathrm{T}}
\textbf{p}\,\mathrm{d}S,
\end{equation}
where $\mathbf{p}=(\tau,\,\sigma,q)^{\mathrm{T}}$. Since the
cohesive traction components $\sigma$ and $\tau$ and the heat flux
$q$ may depend on quantities whose values vary during the
simulation, a consistent linearization of the interface constitutive
law has to be adopted for its use in the Newton-Raphson iterative
method \cite{pawr2}:
\begin{equation}\label{eq9}
\mathbf{p}=\mathbf{C}\,\mathbf{g}=\mathbf{C}\,\mathbf{B}\,\mathbf{R}\,\mathbf{u},
\end{equation}
where the matrix $\mathbf{C}$ is the tangent constitutive matrix of
the element:
\begin{equation}\label{eq10}
\normalsize \mathbf{C}=\left[
\begin{array}{ccc}
\dfrac{\partial \tau}{\partial g_{\mathrm{t}}} & \dfrac{\partial
\tau}{\partial
g_{\mathrm{n}}} & 0\\
\dfrac{\partial\sigma}{\partial g_{\mathrm{t}}} & \dfrac{\partial
\sigma}{\partial g_{\mathrm{n}}} & 0\\
\dfrac{\partial q}{\partial g_{\mathrm{t}}} & \dfrac{\partial
q}{\partial g_{\mathrm{n}}} & \dfrac{\partial
q}{\partial g_T}\\
\end{array}
\right].
\end{equation}
This matrix is in general not symmetric if the Mixed Mode cohesive
zone model has different parameters for the Mode I and the Mode II
traction components. Moreover, examining the coupling with the
thermal field, two off-diagonal terms arise in the third row of
Eq.\eqref{eq10} if the heat conduction constitutive relation is
dependent on the opening and sliding displacements. As we will show
in the sequel, these two terms are equal to zero in the Kapitza
model, which allows for the use of uncoupled schemes and symmetric
solvers.

By introducing Eqs.\eqref{eq7} and \eqref{eq9} into Eq.
\eqref{Gint:1} we get:
\begin{equation}\label{Gint:2}
\delta G_{int}=\delta
\textbf{u}^{\mathrm{T}}\,\textbf{K}\,\textbf{u},
\end{equation}
where
\begin{equation}\label{Gint:3}
\textbf{K}= \textbf{R}^{\mathrm{T}} \int_{S_{int}}
\textbf{B}^{\mathrm{T}}\textbf{C}\textbf{B}\,\mathrm{d}S \textbf{R}
\end{equation}
is the tangent stiffness matrix of the element. In the following
analysis, the integral in Eq. \eqref{Gint:3}, as well as for the
residual vector
\begin{equation}\label{Gint:4}
\textbf{F}= \textbf{R}^{\mathrm{T}} \int_{S_{int}}
\textbf{B}^{\mathrm{T}} \textbf{p}\,\mathrm{d}S,
\end{equation}
will be computed using a two-point Gaussian quadrature scheme. The
heat capacity matrix \textbf{M} is not considered for the interface
element, since it is supposed to have a zero thickness. The
transient regime will be solved according to the backward Euler
method (implicit Euler method), which is a suitable scheme for the
solution of the Fourier heat conduction equation.

This thermoelastic interface element has been implemented as a new
user element in the finite element programme FEAP \cite{feap}.

\section{A thermomechanical cohesive zone model based on microscopical contact relations}
\label{coesivo}

\begin{figure}
\centering
 \includegraphics[width=0.75\textwidth]{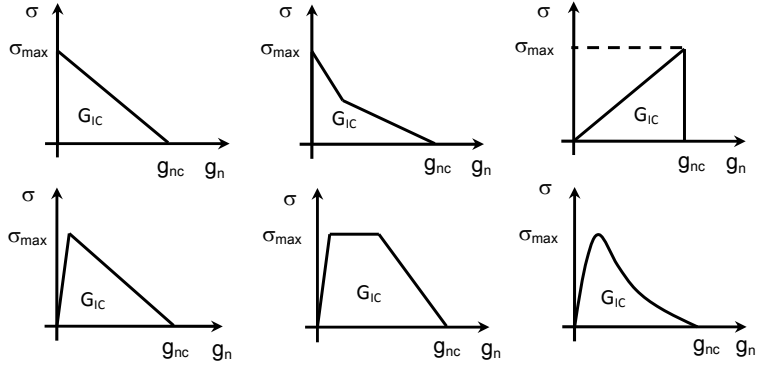}
\caption{Different shapes of the CZMs available in the literature.}
\label{differenti}       
\end{figure}

The progressive separation of an interface due to the propagation of
a crack can be modelled by the cohesive zone model (CZM)
\cite{carp,pawr1}. According to the CZM, a relation between the
normal (Mode I) and tangential (Mode II) cohesive tractions and the
relative opening and sliding displacements experienced by the two
opposite surfaces has to be defined. The various formulations for a
pure Mode I problem are characterized by the peak cohesive traction,
$\sigma_{\max}$, and the Mode I fracture energy, $G_{\mathrm{Ic}}$ ,
which is the area beneath the CZM curve. When the opening
displacement $g_{\mathrm{n}}$ equals a critical value,
$g_{\mathrm{nc}}$, a stress-free crack is created. Different shapes
of the CZM, inspired by atomic potentials, have been proposed so far
(see the qualitative sketch in Fig. \ref{differenti}): linear or
bilinear softening CZMs are usually selected in case of brittle
materials, whereas trapezoidal or bell-shape CZMs are used in case
of ductile fracture. In some cases, linear and bilinear CZMs have an
initial elastic branch with very high stiffness. This branch is
necessary when interface elements are embedded from the very
beginning of the numerical simulation into the finite element mesh
along pre-defined interfaces.

If a suitable relationship between the thermal flux across the crack
faces and the temperature jump is considered, the basic mechanical
CZM formulation can be extended to thermoelastic problems leading to
the so-called thermomechanical CZM. At this point, examining the
state-of-the-art literature, the heat conduction equation was
derived independently of the mechanical part of the CZM
\cite{ozde,hatt,hatt2}. Therefore, additional parameters were
introduced, whose identification is not trivial.

\begin{figure}
\centering
 \includegraphics[width=0.75\textwidth]{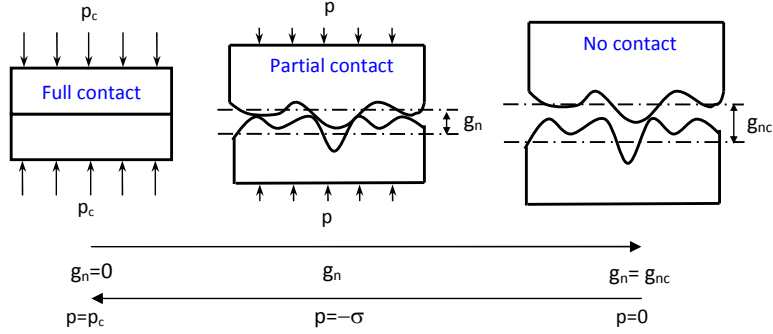}
\caption{Cracked solid loaded by thermal and mechanical loads.}
\label{modello}
\end{figure}

In case of an interface without fibers, an analogy with contact
mechanics can be put forward to simplify the matter. During contact,
a compressive pressure $p$ (negative valued) is applied to the rough
surface and it ranges from zero (first point of contact
corresponding to the tallest asperity) to the full contact pressure,
$p_{\mathrm{c}}$ (see Fig. \ref{modello}). In case of fracture, the
process is basically reversed. The full contact regime can be
regarded as an intact interface and a (positive) tensile traction,
equal in modulus to $p_{\mathrm{c}}$, has to be applied to separate
the two bodies and create a stress-free crack. The process of
debonding progressively produces a rough surface which finally leads
to the microscopically rough stress-free crack (from left to right
in Fig. \ref{modello}). Hence, the Mode I cohesive traction $\sigma$
which, by definition, opposes to crack opening, can be evaluated for
any mean plane separation between the rough surfaces,
$g_{\mathrm{n}}$, as the opposite of the applied contact pressure
$p$ for the same separation.

In case of elastic contact between two bodies with flat or rough
boundaries, a theorem by Barber \cite{barb} demonstrates that the
contact conductance is proportional to the normal contact stiffness.
Hence, taking advantage of this result, it is possible to estimate
the interface contact conductance directly from the solution of the
normal contact problem, without the need of introducing additional
ad hoc constitutive relations for the thermal response. In general,
since the contact stiffness is dependent on the applied pressure,
which is a function of the interface closure, the interface contact
conductance will be dependent on the separation \cite{paba}.
Dimensional analysis considerations and numerical results in
\cite{paba} have demonstrated that the conductance-pressure relation
is of power-law type, with an exponent close to unity. The linear
case is admissible, and has been suggested by Greenwood and
Williamson \cite{GW} with a microscopical contact model which
assumes an exponential distribution of asperity heights.
Independently, the linear relation between conductance and pressure
has been proposed by Persson \cite{persson} in his contact model
rigorously valid in the full contact regime.

Since we are considering a problem of decohesion, where the full
contact regime is the starting point, we consider a linear
conductance-pressure relation as by Greenwood and Williamson and by
Persson and we propose an extension for the application to
decohesion problems. The linear conductance-pressure relation
implies an exponential decay of the cohesive traction w.r.t. the
mean plane separation $g_{\mathrm{n}}$ between the rough surfaces.
In the range $0\leq g_n <l_0$ (very low separations near the full
contact regime), a linear regularization has to be introduced. With
the use of intrinsic interface elements \cite{geubelle} already
embedded in the FE mesh from the beginning of the simulation, an
initial compliance of the CZM is necessary for the equilibrium with
the continuum in the linear elastic regime. Although this
regularization could be regarded as a pure numerical artefact,
actually it can be related to the Young modulus and to the thickness
of the interface region in case of adhesives \cite{allix,pawr3}.
Finally, for separations $g_{\mathrm{n}}$ larger than
$g_{\mathrm{nc}}$, a cut-off to the cohesive tractions corresponding
to the formation of a stress-free crack is introduced. For real
rough surfaces, this cut-off can be set at a distance equal to $3-4$
times the r.m.s. roughness $R$ of the crack profile.

Another modification is needed to consider the weakening effect of
Mixed-Mode deformation, including the effect of the tangential
sliding displacement $g_{\mathrm{t}}$ in the formulation. This can
be done by adding a multiplicative term dependent on
$g_{\mathrm{t}}$ and with the same form as for $g_{\mathrm{n}}$.
According to these modeling assumptions, the resulting expression
for the normal cohesive traction is the following:
\begin{equation}\label{czm1}
\begin{array}{c}
\sigma
\end{array}
\normalsize =\left\{
\begin{array}{lrl}
\sigma_{\max}\,\mathrm{exp}\left(\dfrac{-l_0-|g_{\mathrm{t}}|}{R}\right)\dfrac{g_{\mathrm{n}}}{l_0}, \quad &\mathrm{if}& \quad 0\leq\dfrac{g_{\mathrm{n}}}{R} < \dfrac{l_0}{R}\\
\sigma_{\max}\,\mathrm{exp}\left(\dfrac{-g_{\mathrm{n}}-|g_{\mathrm{t}}|}{R}\right), \quad &\mathrm{if}& \quad \dfrac{l_0}{R}\leq \dfrac{g_{\mathrm{n}}}{R} < \dfrac{g_{\mathrm{nc}}}{R}\\
0, \quad &\mathrm{if}& \quad \dfrac{g_{\mathrm{n}}}{R} \geq \dfrac{g_{\mathrm{nc}}}{R}\\
\end{array}
\right.
\end{equation}

A similar relation can be proposed for the tangential cohesive
tractions:
\begin{equation}\label{czm2}
\begin{array}{c}
\tau
\end{array}
\normalsize =\left\{
\begin{array}{lrl}
\tau_{\max}\,\mathrm{exp}\left(\dfrac{-l_0-g_{\mathrm{t}}}{R}\right)\dfrac{g_{\mathrm{t}}}{l_0}, \quad &\mathrm{if}& \quad 0\leq\dfrac{g_{\mathrm{t}}}{R} < \dfrac{l_0}{R}\\
\tau_{\max}\,\mathrm{sgn}(g_{\mathrm{t}})\,\mathrm{exp}\left(\dfrac{-g_{\mathrm{n}}-|g_{\mathrm{t}}|}{R}\right), \quad &\mathrm{if}& \quad \dfrac{l_0}{R}\leq \dfrac{g_{\mathrm{t}}}{R} < \dfrac{g_{\mathrm{tc}}}{R}\\
 0, \quad &\mathrm{if}& \quad \dfrac{g_{\mathrm{t}}}{R} \geq \dfrac{g_{\mathrm{tc}}}{R}\\
\end{array}
\right.
\end{equation}
where $\tau_{\max}$ and $g_{\mathrm{tc}}$ can be different from
$\sigma_{\max}$ and $g_{\mathrm{nc}}$, respectively.

The interface contact conductance can now be determined via the
first derivative of the normal pressure-separation relation w.r.t.
$g_{\mathrm{n}}$ \cite{barb}:
\begin{equation}\label{czm3}
\begin{array}{c}
k_{\mathrm{int}}
\end{array}
\normalsize =\left\{
\begin{array}{lrl}
\dfrac{1}{\rho_{\mathrm{int}}}, \quad &\mathrm{if}& \quad 0\leq\dfrac{g_{\mathrm{n}}}{R} < \dfrac{l_0}{R}\\
\dfrac{2\sigma}{\rho_{\mathrm{int}}E_{\mathrm{int}}R}, \quad &\mathrm{if}& \quad \dfrac{l_0}{R}\leq \dfrac{g_{\mathrm{n}}}{R} < \dfrac{g_{\mathrm{nc}}}{R}\\
0, \quad &\mathrm{if}& \quad \dfrac{g_{\mathrm{n}}}{R} \geq \dfrac{g_{\mathrm{nc}}}{R},\\
\end{array}
\right.
\end{equation}
where a dependency on the normal contact pressure comes into play in
the range $l_0\leq g_{\mathrm{n}}<g_{\mathrm{nc}}$. The resistivity
$\rho_{\mathrm{int}}$ and the Young modulus $E_{\mathrm{int}}$ of
the interface can be evaluated as
$\rho_{\mathrm{int}}=\rho_++\rho_-$ and
$E_{\mathrm{int}}=[(1-\nu_{+}^2)/E_{+}+(1-\nu_{-}^2)/E_{-}]^{-1}$,
where the subscripts $-$ and $+$ refer to the two materials
separated by the interface and $\nu$ is the Poisson's ratio. In the
range $0\leq g_n <l_0$, a constant interface conductivity is
selected. Since the maximum interface conductivity in contact
problems can be attained for a small separation larger than zero,
the parameter $l_0$ can be chosen according to this physical
argument. In this range, with a constant
$k_{\mathrm{int}}=1/\rho_{\mathrm{int}}$, the present approach is
equivalent to the Kapitza model, where the coefficient
$\rho_{\mathrm{int}}$ should be regarded as the Kapitza resistance.

As compared to previous thermomechanical CZMs, the main advantage of
the proposed formulation relies in the fact that the thermal part of
the CZM is simply derived from the normal stiffness and therefore it
does not introduce additional independent model parameters. The
complete thermomechanical CZM is therefore fully defined in terms of
the maximum (peak) normal and shear tractions $\sigma_{\max}$ and
$\tau_{\max}$, the critical gaps $g_{\mathrm{nc}}$ and
$g_{\mathrm{tc}}$, the r.m.s. roughness $R$, an internal length
$l_0$, the composite thermal resistance $\rho_{\mathrm{int}}$ and
the composite Young's modulus $E_{\mathrm{int}}$. Parameter
identification should be carried out by choosing $\sigma_{\max}$ and
$\tau_{\max}$ to capture the peak stresses deduced from tensile and
shear tests on representative volume elements. The parameters
$g_{\mathrm{nc}}$ and $g_{\mathrm{tc}}$ should be chosen to match
the fracture toughness of the material. The additional parameter
$l_0$ should be selected according to the physical compliance of the
interface as proposed in \cite{pawr3}. The r.m.s. roughness $R$ can
be quantified from a profilometric analysis of the crack path at
failure. Finally, $\rho_{\mathrm{int}}$ has to be related to the
resistivities of the bulk materials and it should be equal to the
Kapitza resistance.

\begin{figure}[h!]
\centering \subfigure[Mode I
CZM]{\includegraphics[width=.45\textwidth,angle=0]{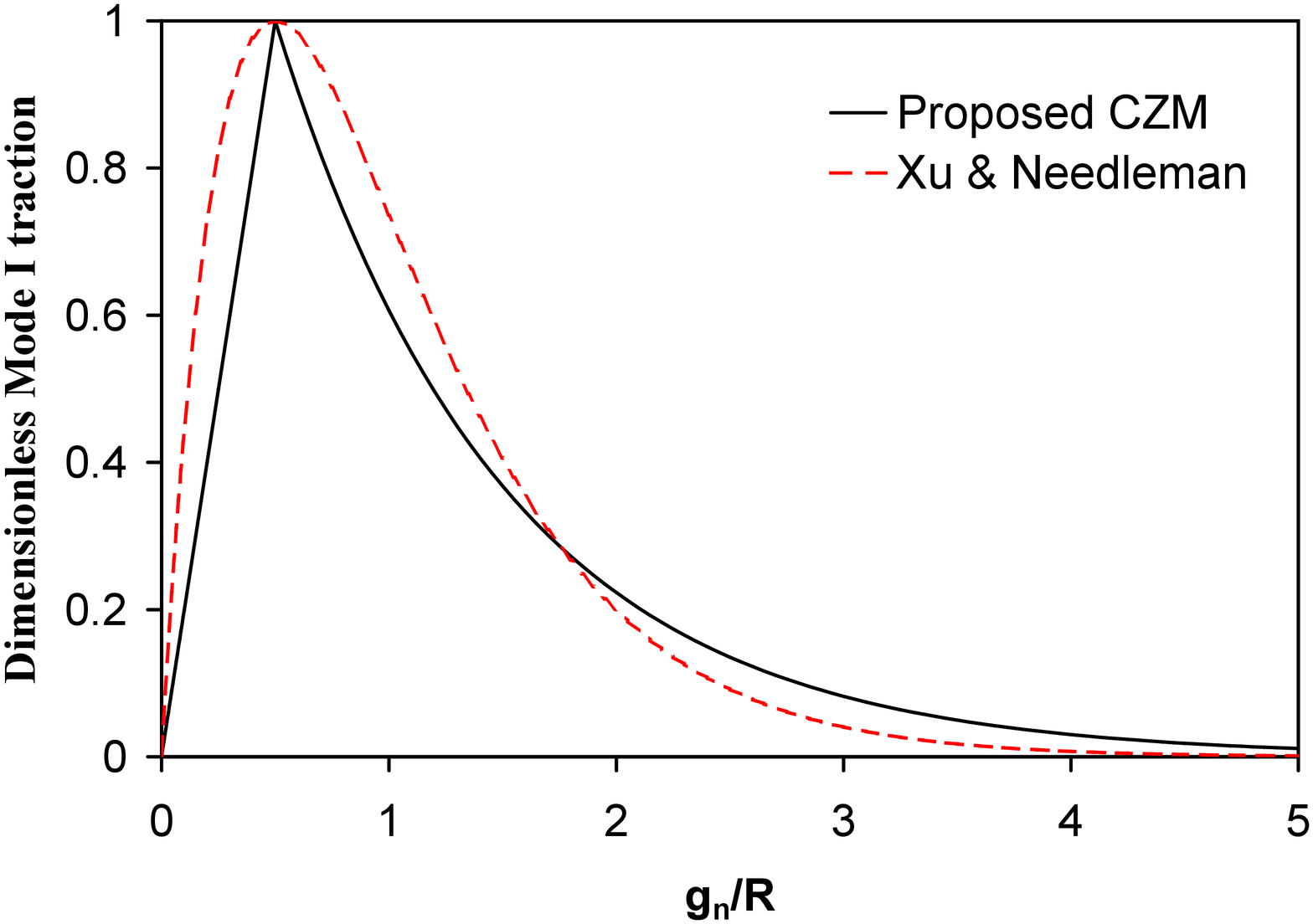}\label{czmxu_a}}\\
      \subfigure[Interface conductance]{\includegraphics[width=.45\textwidth,angle=0]{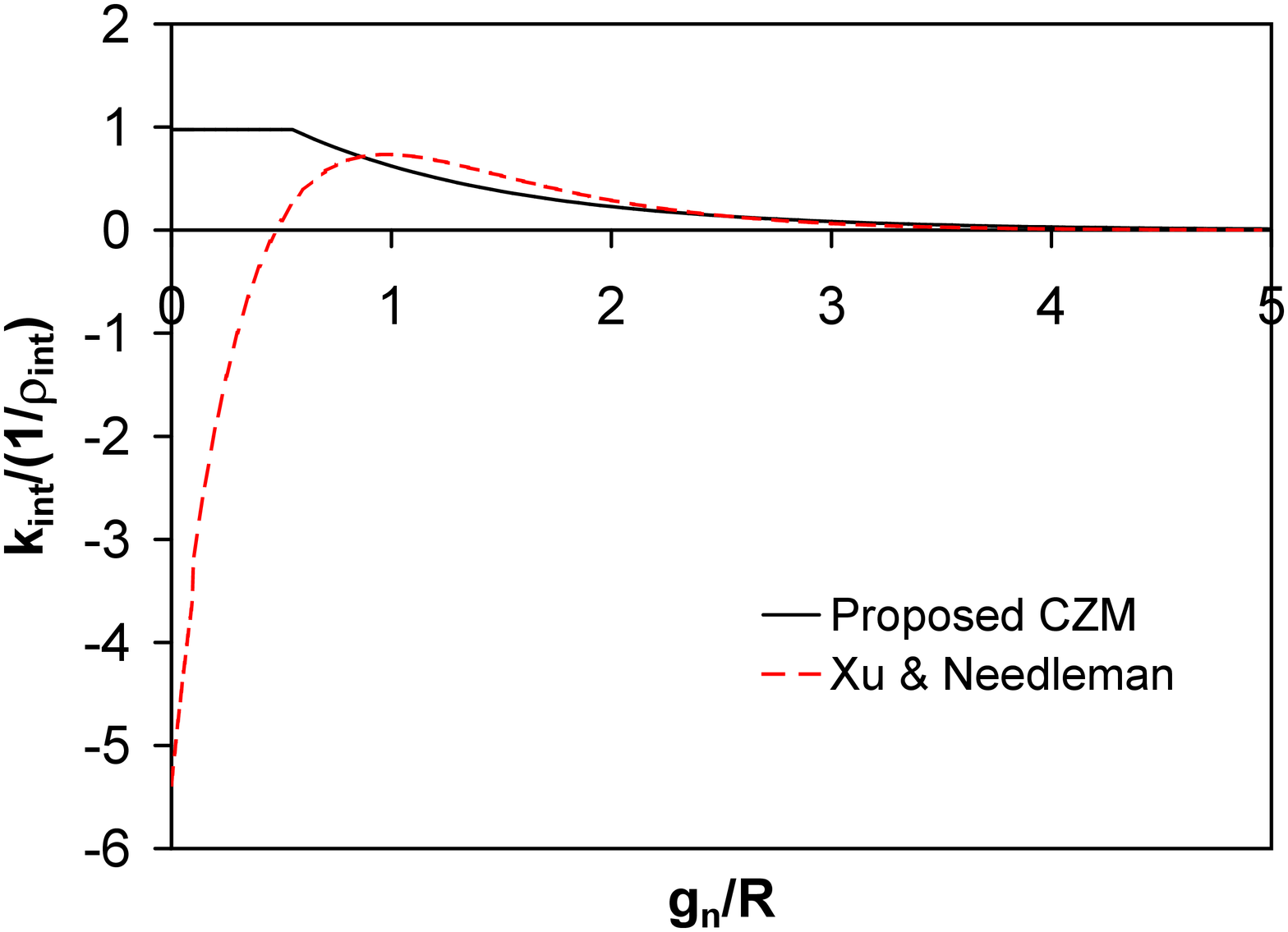}\label{czmxu_b}}
\caption{Comparison between the proposed CZM and that by Xu and
Needleman \cite{xu} with matched parameters.}\label{czmxu}
\end{figure}

The normal cohesive traction \eqref{czm1} is plotted vs.
$g_{\mathrm{n}}$ in Fig.\ref{czmxu}. It is interesting to note the
similitude between the present formulation deduced according to
contact mechanics considerations and the CZM by Xu and Needleman
\cite{xu} and its subsequent generalizations \cite{bosch}. In
\cite{xu}, the shape of the CZM is the result of the product between
a linear function of the gap (dominating for small separations) and
an exponential decay (prevailing for large separations), see the
dashed curve in Fig.\ref{czmxu_a}. Although the shape of the Xu and
Needleman CZM is not so different from the proposed expression and
has the advantage of being defined by a single equation for the
whole range of separations, if we attempt at estimating the
interface contact conductance by differentiating it w.r.t.
$g_{\mathrm{n}}$ we obtain an unphysical result. The interface
contact conductance is negative at the beginning and it approaches
that predicted by the present model only for very large separations,
see Fig.\ref{czmxu_b} obtained from the curves in Fig.\ref{czmxu_a}.

As a result of the proposed model, the interface conductance
\eqref{czm3} does depend on the separation. The heat flux normal to
the interface is given by $q=-k_{\mathrm{int}}\,g_T$. Its consistent
linearization according to Eq.\eqref{eq10} provides the following
terms for $0\leq g_{\mathrm{n}}/R<l_0/R$:
\begin{subequations}
\begin{align}
\dfrac{\partial q}{\partial g_{\mathrm{t}}}&=0,\\
\dfrac{\partial q}{\partial g_{\mathrm{n}}}&=0,\\
\dfrac{\partial q}{\partial g_T}&=-\dfrac{1}{\rho_{\mathrm{int}}},
\end{align}
\end{subequations}
and for $g_{\mathrm{n}}/R<g_{\mathrm{nc}}/R$:
\begin{subequations}
\begin{align}
\dfrac{\partial q}{\partial g_{\mathrm{t}}}&=-\dfrac{2\, g_T}{\rho_{\mathrm{int}}\,E_{\mathrm{int}}\,R}\dfrac{\partial\sigma}{\partial g_{\mathrm{t}}},\\
\dfrac{\partial q}{\partial g_{\mathrm{n}}}&=-\dfrac{2\, g_T}{\rho_{\mathrm{int}}\,E_{\mathrm{int}}\,R}\dfrac{\partial\sigma}{\partial g_{\mathrm{n}}},\\
\dfrac{\partial q}{\partial
g_T}&=-\dfrac{2\sigma}{\rho_{\mathrm{int}}\,E_{\mathrm{int}}\,R}.
\end{align}
\end{subequations}

\section{Numerical results}

In this section we propose a simple example where we compare the
present CZM predictions with those based on the Kapitza constant
resistance model.

\begin{figure}
\centering
  \includegraphics[width=0.45\textwidth]{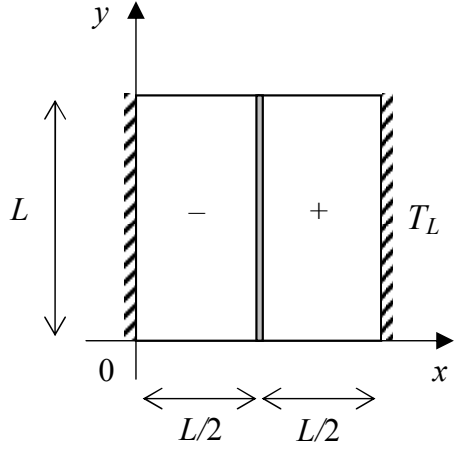}
\caption{Square domain with a cohesive crack.} \label{geometria}
\end{figure}

A bi-material composite of lateral side $L$, clamped at $x=0$ and at
$x=L$ is considered (Fig. \ref{geometria}). A cohesive interface is
placed at $x=L/2$. For the sake of simplicity, the bodies are
assumed to have identical material properties.

An initial temperature $T_i$ is prescribed over the whole bodies and
a temperature $T_L$ $(T_L < T_i)$ is imposed along the right side
($x=L$, Fig. \ref{geometria}). We let the temperature vary inside
the two bodies and along the other boundaries.  Due to cooling of
the right hand side, the material region $+$ will shrink more than
the region $-$ and will progressively put in tension the interface
until a possible debonding. This could be the case of a building
wall with exposed surface on the right side.

According to dimensional analysis arguments, for a given $y$, once
the parameters $\nu$, $l_0$ and $R$ are prescribed, the temperature
field $T$ throughout the body is a function of nine parameters:
\begin{equation}\label{para1}
T=T(x,L,T_i,T_L,t,D,E,g_{\mathrm{nc}},\sigma_{\max}),
  \end{equation}
where $D=k/(d_m c)$ is the thermal diffusion coefficient of the
bulk. A dimensionless temperature $T^*=(T-T_i)/(T_i-T_L)$ can be
introduced for the analysis of the results. A straightforward
application of Buckingham theorem allows us reducing the dependency
of $T^*$ on four parameters:
\begin{equation}\label{para2}
T^*=T^*\left(x^*,t^*,g_{\mathrm{nc}}^*,\sigma_{\max}^*\right),
\end{equation}
where:
\begin{equation*}
x^*=\dfrac{x}{L},\; t^*=\dfrac{t D}{L^2},\;
g_{\mathrm{nc}}^*=\dfrac{g_{\mathrm{nc}}}{L},\;
\sigma_{\max}^*=\dfrac{\sigma_{\max}}{E}.
\end{equation*}
Notice that, when dealing with Kapitza's model, also the interface
dimensionless conductivity $k^*=k/k_{\mathrm{int}}$ will be
introduced (Eq.\eqref{heat1:2}), for the sake of clarity.

In the next sections, numerical predictions will refer to $y/L=0.5$
by assuming plane stress conditions. The following parameters will
be selected: $\nu=0.1$, $l_0/R=0.01$, $\sigma_{\max}^*=0.032$ and
$g_{\mathrm{nc}}^*=0.05$.

\subsection{Predictions according to Kapitza model}\label{kap5}

To provide a reference solution for quantifying the role of
thermomechanical coupling in the interface constitutive relations,
we first consider the simplified Kapitza model where the interface
conductivity is a constant value. Hence,
$k_{\mathrm{int}}=\mathrm{const}$ and this is the only non vanishing
term entering the tangent constitutive matrix \eqref{eq10}. The
partial derivatives of the heat flux with respect to the normal and
tangential gaps are zero. Therefore, the heat conduction equation
and the equations of equilibrium become uncoupled in this case. This
simplification allows for the implementation in FE codes where the
mechanical and the thermal fields are solved separately. In
particular, the thermal field should be solved first. Afterwards,
the thermoelastic deformation has to be computed by solving the
mechanical problem. The evolution of debonding will depend on the
normal and tangential gaps at the interface and the stress field in
the horizontal direction will be imposed by the mechanical part of
the CZM constitutive relations.

\begin{figure}
\centering
  \includegraphics[width=0.6\textwidth]{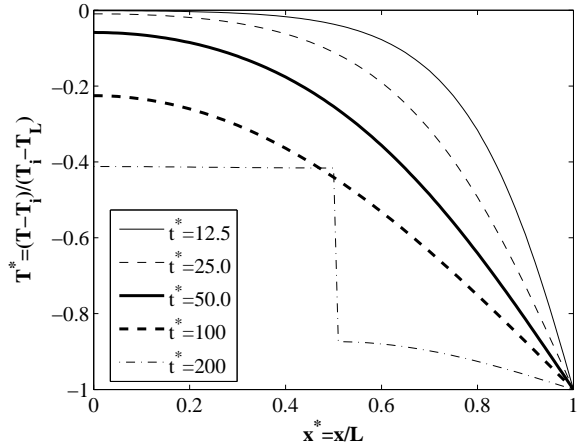}
\caption{Kapitza's model, $k^*=0.001$: dimensionless temperature
field vs. dimensionless time.} \label{kap1}
\end{figure}

Three cases are examined depending on the value of $k^*$, i.e.,
$k^*=0.001$, $k^*=1$ and $k^*=1000$. In Fig.\ref{kap1}, $k^*=0.001$
and the interface is highly conductive. A parabolic profile of the
temperature, with no discontinuities, is initially observed along
$x^*$. At $t^*\simeq 125$, the interface debonds and the interface
conductivity suddenly jumps from the value of $k^*$ to zero.
Increasing time, the temperature of the two half bodies stabilize:
the dimensionless temperature of the right part is progressively
decreasing with time $t^*$ down to $-1$ .

\begin{figure}
\centering
  \includegraphics[width=0.6\textwidth]{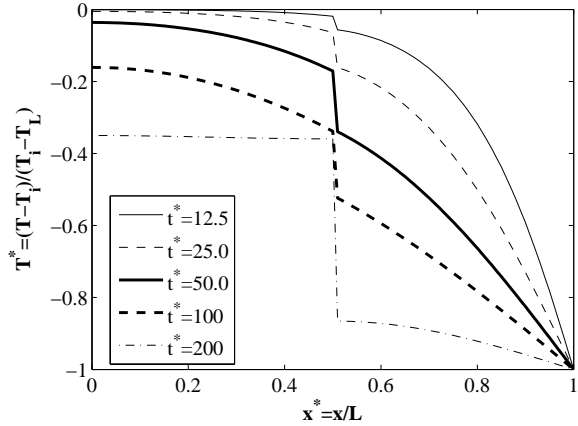}
\caption{Kapitza's model, $k^*=1$: dimensionless temperature field
vs. dimensionless time.} \label{kap2}
\end{figure}

For $k^*=1$ (see Fig.\ref{kap2}), a temperature discontinuity is
observed across the interface, since it is no longer highly
conductive and it imposes a localized additional resistance to the
system. Debonding takes place at $t^*\simeq 130$, similarly to the
previous case.

\begin{figure}
\centering
  \includegraphics[width=0.6\textwidth]{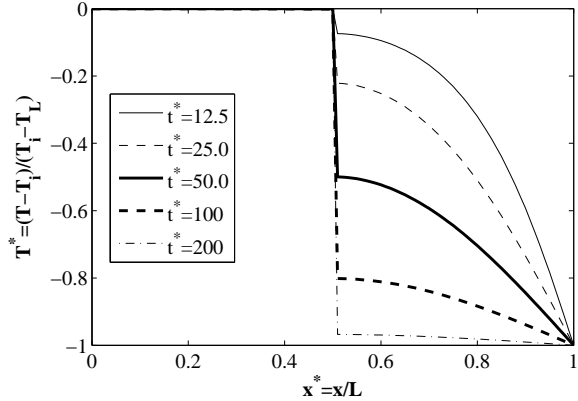}
\caption{Kapitza's model, $k^*=1000$: dimensionless temperature
field vs. dimensionless time.} \label{kap3}
\end{figure}

For $k^*=1000$ (see Fig.\ref{kap3}), the interface plays the role of
an insulator and only the temperature of the right hand side varies
with time, tending to the imposed value of $-1$. A negligible heat
flux enters the left hand side, whose temperature remains nearly
constant and equal to the initial one. For the chosen CZM parameters
and the imposed dimensionless temperature jump, debonding does not
take place in this case.

\subsection{Proposed CZM predictions}

The coupled thermomechanical CZM predictions are now presented. In
this case, all the terms entering the tangent constitutive matrix
\eqref{eq10} are different from zero and are considered. The
solution is gained by solving for the thermal and the mechanical
fields at the same time, since the crack opening influences the
interface conductivity and therefore the solution of the thermal
field. Moreover, an unsymmetrical solver is used due to the non
symmetry of the interface element stiffness matrix.

\begin{figure}
\centering
\includegraphics[width=0.6\textwidth]{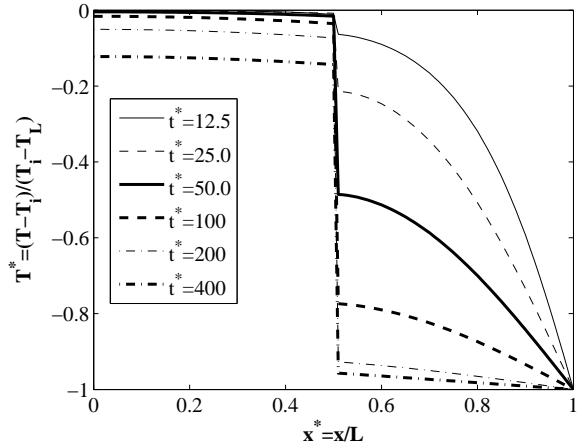}
\caption{CZM predictions, dimensionless temperature field vs.
dimensionless time.} \label{tempo1}
\end{figure}

\begin{figure}
\centering
\includegraphics[width=0.6\textwidth]{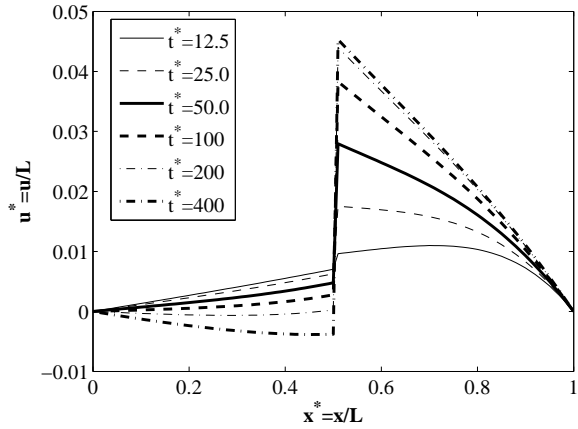}
\caption{CZM predictions, dimensionless displacement field vs.
dimensionless time.} \label{tempo2}
\end{figure}

The temperature and the horizontal displacement distributions
predicted by the proposed CZM are shown in Fig.\ref{tempo1} and
\ref{tempo2}, respectively, for different dimensionless times $t^*$.

Examining Fig.\ref{tempo1}, the temperature jump at the interface is
initially an increasing function of $t^*$ due to the cooling of the
right-hand side. Then the jump decreases: this process is relatively
slow, due to the progressive opening of the cohesive crack which
reduces the interface conductivity. Debonding takes place for
$t^*\simeq 400$.

Looking at the displacement field (Fig.\ref{tempo2}), two different
stages are observed in the transient regime. In the range $0<t^*\leq
12.5$, the temperature of the right part decreases and the body
progressively shrinks (positive displacements, i.e., displacements
directed to the right). Due to the cohesive tractions transmitted by
the interface, whose dilatation effect initially overcomes the
thermal contraction in the left part of the body, a net positive
displacement is observed for $x^*<0.5$. For $t^*>12.5$, the cohesive
tractions reduce in magnitude due to the increased normal gap
(softening regime) and the thermal contraction effect prevails. As a
result, the left part experiences negative displacements, i.e.,
leftward.

\begin{figure}
\centering
\includegraphics[width=0.6\textwidth]{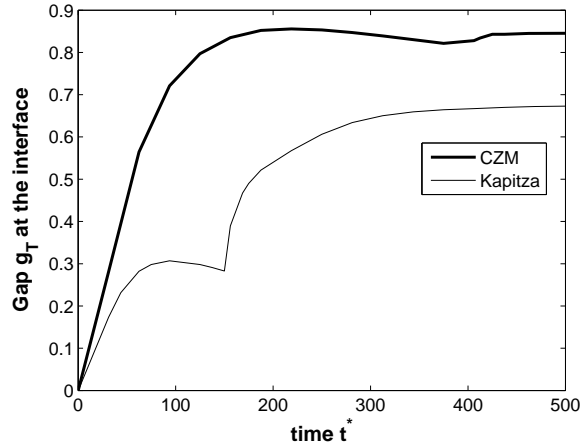}
\caption{Temperature gap $g_T$ at the interface: comparison between
CZM and Kapitza model.} \label{paggi1}
\end{figure}

\begin{figure}
\centering
\includegraphics[width=0.6\textwidth]{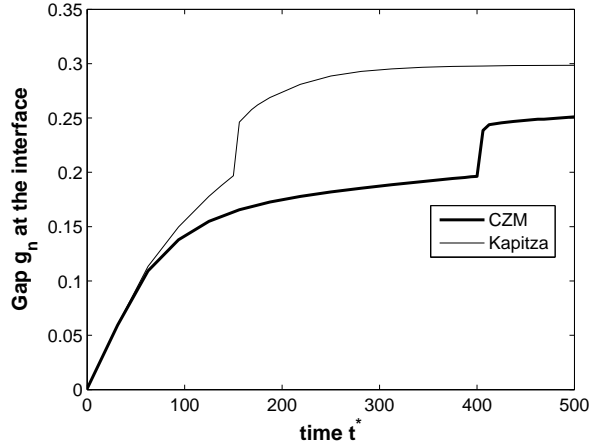}
\caption{Normal gap $g_{\mathrm{n}}$ at the interface: comparison
between CZM and Kapitza model.} \label{paggi2}
\end{figure}

A closer comparison with the Kapitza model can be made by comparing
the CZM predictions with those predicted by the Kapitza model with
the same $k_{\mathrm{int}}$. The absolute temperature gap $g_T$ and
the normal displacement gap $g_{\mathrm{n}}$ at the interface are
shown in Figs.\ref{paggi1} and \ref{paggi2}, respectively, as
functions of time $t^*$.

At the very beginning of the simulation, for $g_n/R<l_0/R$, the
proposed thermoelastic CZM and the Kapitza model provide the same
response. Later on, the predictions of the two models diverge, due
to the reduction of interface conductance related to the increased
normal gap in the thermoelastic CZM.

As already observed, the thermal gap predicted by the proposed
thermoelastic CZM rapidly rises. Later on, it decreases slowly until
debonding takes place for $t^*\simeq 400$, where a small
discontinuity in $g_T$ is observed. The Kapitza model presents a
similar trend, but debonding takes place much earlier, for
$t^*\simeq 130$.

The interface normal gap, $g_{\mathrm{n}}$ is shown in Fig.
\ref{paggi2}. After an initial matching (imputable to combined
thermoelastic effects, holding also for $g_{\mathrm{n}}>l_0$ where
the two constitutive models are different), the crack opening
predicted by the proposed CZM is smaller than that by the Kapitza
model, due to the reduced interface conduction. Debonding takes
place at the same $g_{\mathrm{nc}}$, since the mechanical part of
the CZM is the same for both simulations, but for very different
times.

\begin{figure}
\centering
\includegraphics[width=0.6\textwidth]{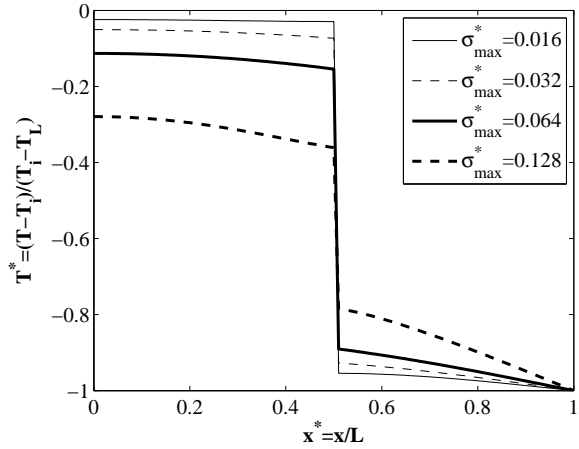}
\caption{The effect of $\sigma_{\max}^*$ on the temperature field
predicted by the proposed CZM.} \label{sigma1}
\end{figure}

\begin{figure}
\centering
\includegraphics[width=0.6\textwidth]{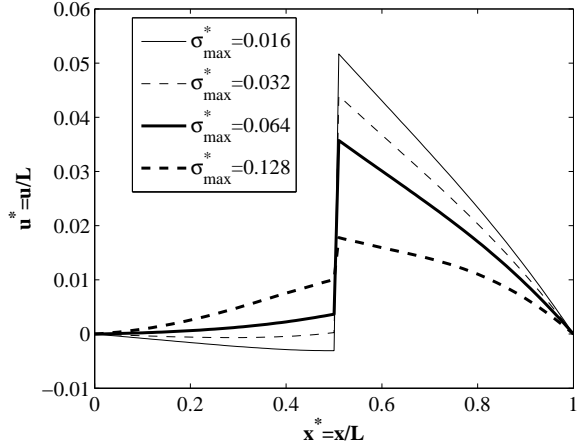}
\caption{The effect of $\sigma_{\max}^*$ on the horizontal
displacement field predicted by the proposed CZM.} \label{sigma2}
\end{figure}

Finally, the effect of the CZM parameter $\sigma_{\max}^*$
($g_{\mathrm{nc}}^*=0.05$) is shown in Figs.\ref{sigma1} and
\ref{sigma2} for $t^*=200$. By reducing $\sigma^*_{\max}$, debonding
takes place earlier. This is due to the competition between the
strain inducted by the mechanical CZM tractions and the shrinkage
due to thermal strains. For small values of $\sigma^*_{\max}$, the
net displacement in the left part of the body is negative (thermal
strain prevailing over the mechanical one) and the normal gap is
amplified. For large values of $\sigma^*_{\max}$, the opposite
situation takes place, the horizontal displacement is positive
everywhere and the normal gap is reduced.

A similar trend is observed by varying $g_{\mathrm{nc}}^*$ keeping
fixed $\sigma_{\max}^*=0.032$.

\subsection{Effect of gas conductivity}

\begin{figure}
\centering
\includegraphics[width=0.75\textwidth]{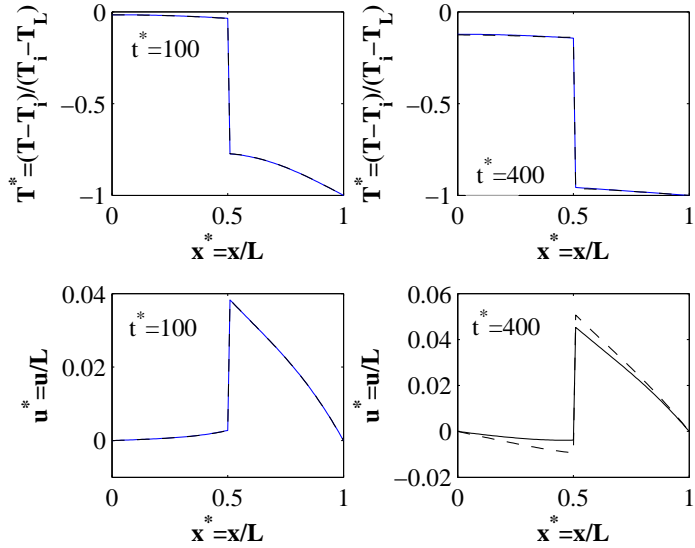}
\caption{Effect of the gas conductivity (dashed line) on the
temperature and displacement fields. Predictions without gas
conductivity are shown with continuous line.} \label{gas2}
\end{figure}

In previous studies \cite{ozde,hatt}, the contribution of the gas to
the interfacial conductivity was found to be significant. Clearly,
this might depend on the problem at hand and no general rules can be
put forward. To assess its effect for the present CZM formulation,
the gas contribution can be included as follows:
\begin{equation}
q=-k_{\mathrm{int}}\Delta T+k_{\mathrm{gas}},
\end{equation}
where $k_{\mathrm{gas}}$ is the gas conductivity. In the present
work, we assume $k_{\mathrm{gas}}=k/1000$ as in \cite{ozde,hatt}.
Results are shown in Fig.\ref{gas2}. The gas contribution to the
conductivity is significant only for significant crack openings and
in general for $t^*\geq 400$. The gas conductivity has a negligible
influence on the time for fracture initiation.

\section{Application to photovoltaics}

The present research on thermomechanical CZM is the continuation of
the previous study in \cite{casa1}, where a multi-scale and
multi-physics computational approach has been proposed to
investigate the effect of cracking in silicon used for solar cells.
Experimental results in \cite{WSZ} have shown that the electric
conductivity of cracks is highly dependent on the temperature field.
This effect is attributed both to the physics of the semiconductor
whose governing equations strongly depend on the temperature, and by
possible self-healing of cracks due to closing induced by
thermoelastic deformation \cite{casa}.

\begin{figure}
\centering
\includegraphics[width=0.6\textwidth]{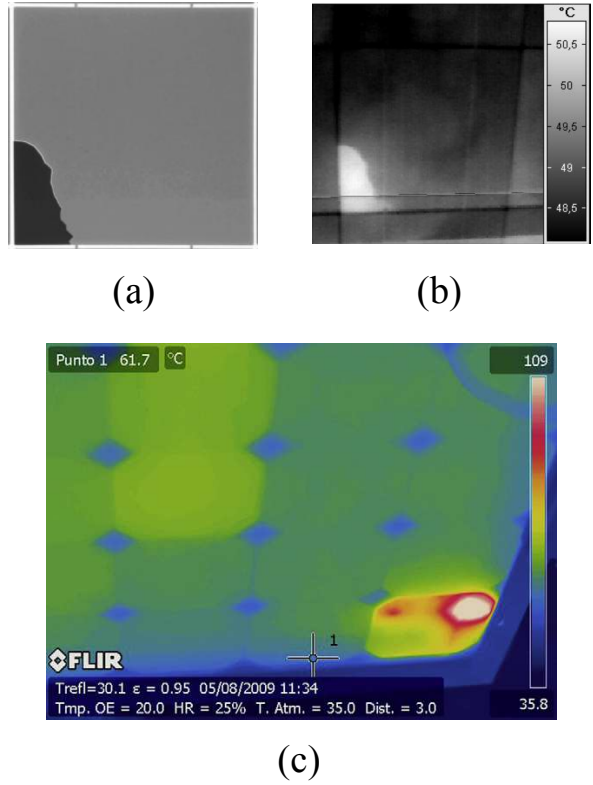}
\caption{Thermal images obtained with a thermocamera showing local
temperature rises (hot spots) in silicon cells in case of cracks
((a) and (b) adapted from \cite{WSZ}, (c) from
\cite{thermocamera2}).} \label{fig17}
\end{figure}

Examining the problem in more details, we know that during the
production of a photovoltaic module crack-free cells made of mono-
or polycrystalline silicon are laminated inside a stack composed of
an encapsulating polymer and a cover glass at a temperature of about
$T_0=150$ $^\circ$C. Later on, the module is brought to the
environmental temperature and cracks can be inserted by handling,
transport and installation operations. In proximity of a crack, the
local temperature can rise significantly, leading to the so called
\emph{hot spot}, as evidenced in the thermal images of
Fig.\ref{fig17}.

\begin{figure}[h!]
\centering \subfigure[Sketch of the domains separated by a
crack]{\includegraphics[width=.4\textwidth,angle=0]{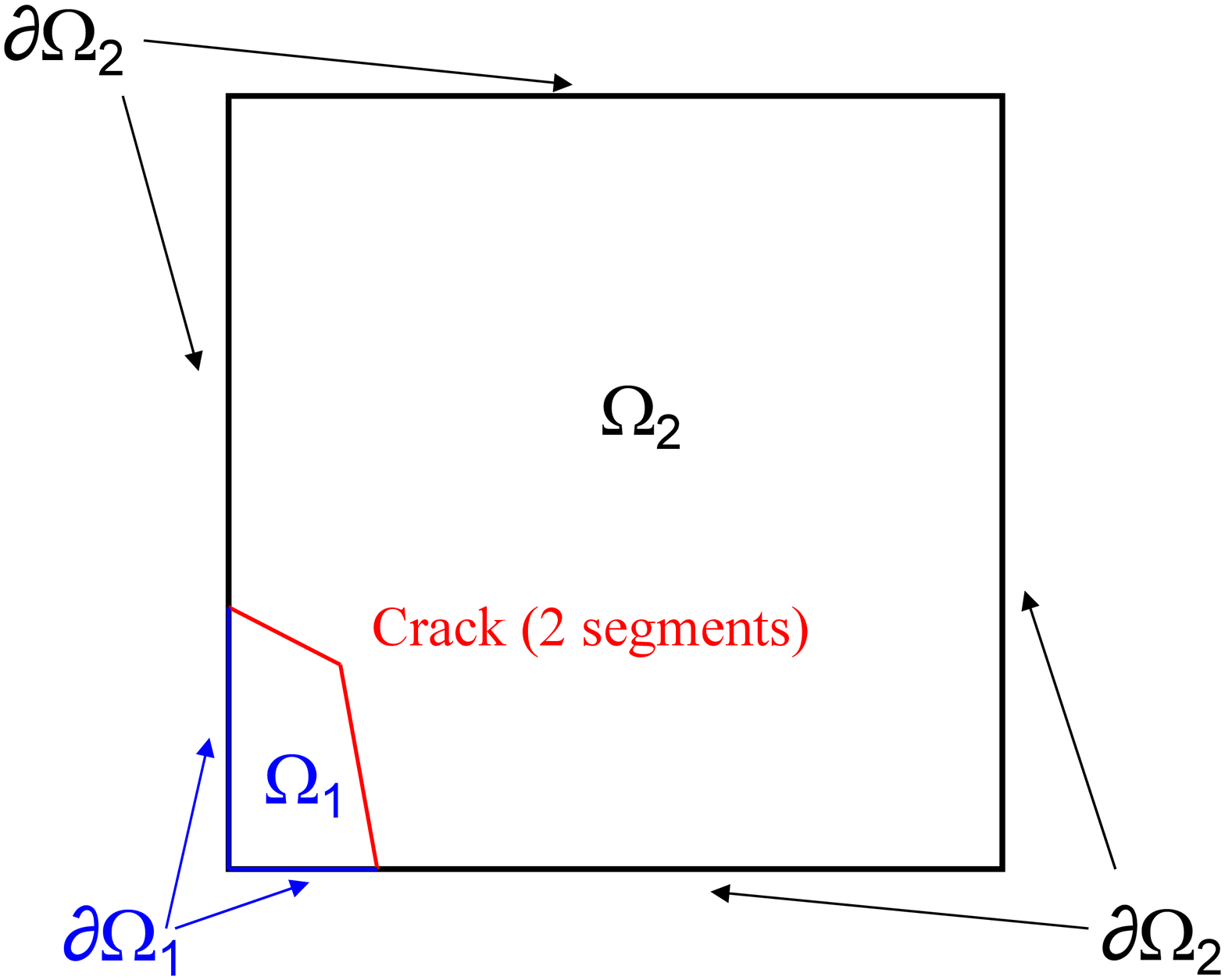}\label{fig18a}}\quad
      \subfigure[FE mesh $(n=8)$]{\includegraphics[width=.32\textwidth,angle=0]{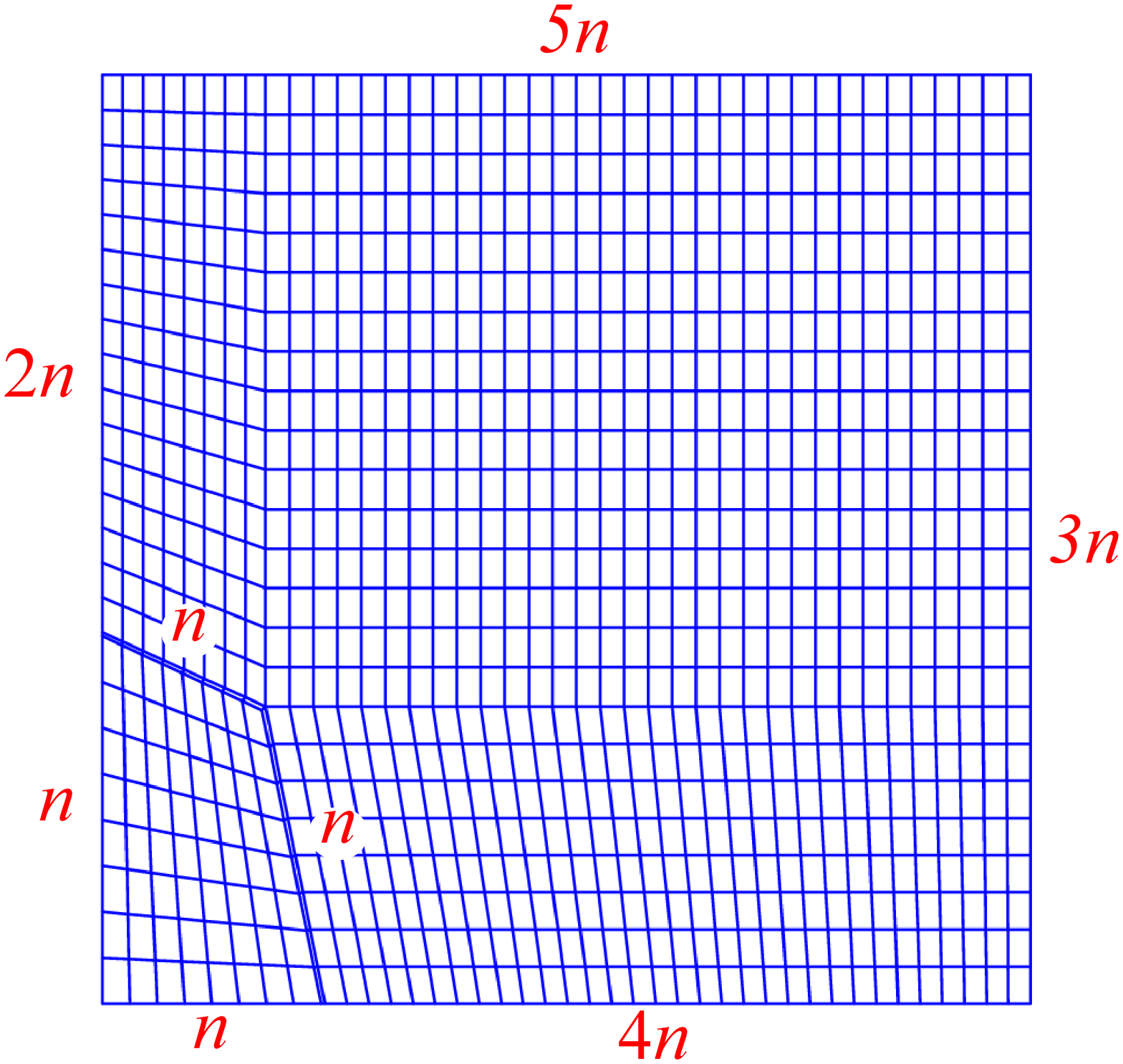}\label{fig18b}}
\caption{(a) Sketch of the geometry of a solar cell with a crack in
a corner. Temperature excursions $\Delta T_1=-100$ $^\circ$C and
$\Delta T_2=-120$ $^\circ$C from the reference stress-free
temperature $T_0=150$ $^\circ$C are imposed along external
boundaries $\partial\Omega_1$ and $\partial\Omega_2$, respectively.
(b) FE mesh generated with block commands in FEAP. Different
discretizations can be achieved by varying $n$.}\label{fig18}
\end{figure}

As a model example, we consider a solar cell made of monocrystalline
silicon with a crack on one corner, see Fig.\ref{fig18}, similar to
the real case shown in Fig.\ref{fig17}(a). The crack is modelled by
inserting interface elements along the two crack segments. The
vertical sides of the cell are constrained to displacements in the
horizontal direction, whereas the vertical sides are constrained in
the vertical direction.

An initial temperature is applied to the cell boundaries. According
to Fig.\ref{fig18a}, the crack separates the cell in two domains, a
small one potentially insulated from the electrical point of view,
$\Omega_1$, and the rest of the undamaged cell, $\Omega_2$. The
whole external boundary, $\partial\Omega$, can also be partitioned
into two parts: $\partial\Omega_1$ and $\partial\Omega_2$. On
$\partial\Omega_1$, an initial temperature excursion $\Delta
T_1=-100$ $^\circ$C from the stress-free state at $T=T_0$ is
imposed, which corresponds to the jump from the lamination
temperature to an operating temperature of $50$ $^\circ$C, to
simulate the presence of a hot spot. On $\partial\Omega_2$, we set
$\Delta T_2=-120$ $^\circ$C, i.e., a lower operating temperature of
30$^\circ$C. Different FE meshes are considered by varying the
parameter $n$, see Fig.\ref{fig18b}.

The underlying nonlinear transient heat conduction problem is solved
in order to determine the temperature distribution in the solar cell
vs. time. According to the symbology introduced in Section 2, the
material parameters of Silicon are $E=169$ GPa, $\nu=0.16$,
$d_m=3100$ kg/m$^3$, $k=114$ W/(m$^\circ$C), $c=715$
J/(kg$^\circ$C). The coefficient of thermal expansion is
$\alpha=1.1\times 10^{-6}$ 1/C$^\circ$. The cell thickness is
$0.166$ mm. Regarding the CZM, we simulate a material with a tensile
strength of about $1$ GPa, in the range of typical values reported
for Silicon. The fracture energy is $G_{\mathrm{IC}}=5.92$ N/m. From
this toughness and the functional form of the CZM we can deduce the
values of the remaining parameters:
$g_{\mathrm{nc}}=0.2\,\mu\mathrm{m}$, $R=3.135\times
10^{-2}\,\mu\mathrm{m}$, and $l_0=3.135\times
10^{-4}\,\mu\mathrm{m}$.

\begin{figure}
\centering
\includegraphics[width=0.6\textwidth]{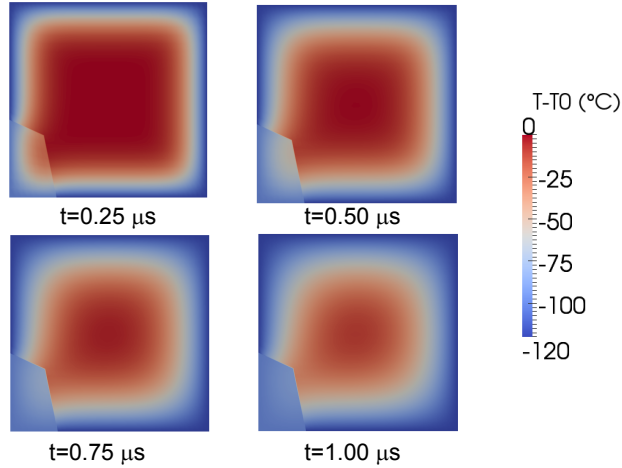}
\caption{Time evolution of the temperature field in the solar cell
with a crack in the corner. The reference stress-free temperature is
$T_0=150$ $^\circ$C.} \label{fig19}
\end{figure}

The temperature field in the solar cell is shown in Fig.\ref{fig19}
for a sequence of times. The region of the cell in the corner,
separated by the crack, tends very rapidly to a uniform temperature
equal to that imposed along its boundary.

\begin{figure}
\centering
\includegraphics[width=0.8\textwidth]{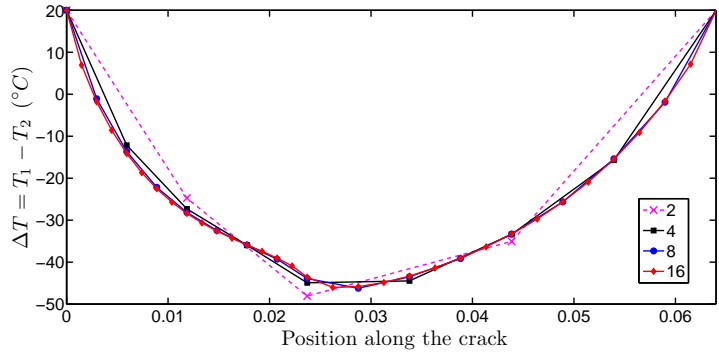}
\caption{Mesh convergence study: temperature jump $T_1-T_2$ across
the crack faces vs. position along the crack for different number of
interface elements $n$ used to discretize the crack segments.}
\label{fig20}
\end{figure}

The result of a mesh convergence study by varying the parameter $n$
setting the number of interface elements per crack segment from 2 to
16 (see an example in Fig.\ref{fig18b} for $n=8$) is shown in
Fig.\ref{fig20}. It depicts the temperature jump across the crack
faces between region 1 (warmer) and region 2 (cooler) vs. a
curvilinear coordinate moving along the two crack segments and
starting from the emergent point of the crack on the vertical left
side. FE solutions by varying $n$ converge very fast and the
discrepancy between the solutions for $n=8$ and $n=16$ elements per
crack segment is almost negligible.

\section{Conclusions}
A coupled thermo-mechanical CZM derived according to an analogy with
contact mechanics between rough surfaces has been proposed: the
crack conductivity results to be a function of the normal cohesive
tractions and the model captures the transition from the Kapitza
constant resistance approach, valid for a negligible crack opening,
to a crack-opening interface conductivity in case of partial
debonding. Thermo-elastic effects related to the transient regime
have been investigated, with particular attention to: (i) the time
evolution of the temperature and displacements fields; (ii) the
influence of the cohesive parameters on fracture initiation; (iii)
the influence of the gas conductivity. It has also been evidenced
that, neglecting thermoelastic coupling, as assumed by Kapitza's
model, very different thermal and mechanical responses are obtained.
Therefore, the application of the Kapitza model to thermomechanical
configurations where the phenomenon of interface debonding may occur
should be checked with care.

An application to photovoltaics has been finally provided, showing
the potentiality of the method to model the transient regime in the
thermoelastic field in bodies containing cohesive cracks. Future
perspectives of this work regard the further coupling with the
electric field which, according to the physics of the solar cell,
takes place in the direction orthogonal to the surface of the solar
cell and is significantly influenced by cracks and defects.

\vspace{1em} \addcontentsline{toc}{section}{Acknowledgements}
\noindent\textbf{Acknowledgements} \vspace{1em}

The research leading to these results has received funding from the
European Research Council under the European Union's Seventh
Framework Programme (FP/2007-2013)/ERC Grant Agreement No. 306622
(ERC Starting Grant Multi-field and multi-scale Computational
Approach to Design and Durability of PhotoVoltaic Modules - CA2PVM).
The support of the Italian Ministry of Education, University and
Research to the Project FIRB 2010 Future in Research Structural
mechanics models for renewable energy applications (RBFR107AKG) is
gratefully acknowledged.

\end{document}